\documentclass[prl,superscriptaddress,floatfix,showpacs,aps,reprint,letterpaper,nobalancelastpage]{revtex4-1}

\usepackage{graphicx}
\usepackage{color} 
\usepackage{amsthm}
\usepackage{amsmath}
\usepackage{amssymb}
\usepackage{bbm,amsfonts}
\usepackage{setspace}
\usepackage{nomencl}
\usepackage{amstext}

\newcommand{\Lam}{\Lambda}
\newcommand{\Id}{\mathbbm{1}}

\newcommand{\Paulone}{\mathcal{P}_1}
\newcommand{\Pauln}{\mathcal{P}_n}

\newcommand{\braa}{\langle \psi |}
\newcommand{\kett}{|\psi\rangle}

\newcommand{\Hilb}{\mathcal{H}}

\newcommand{\Clif}{\text{Clif}_n}
\newcommand{\Clifone}{\text{Clif}_1}

\newcommand{\tr}{{\mathrm{tr}}}

\newcommand{\LamC}{\Lambda_{\mathcal{C}}}

\newcommand{\LamP}{\Lambda_{\mathcal{P}}}
\newcommand{\LamD}{\Lambda_{\mathcal{D}}}
\newcommand{\LamA}{\Lambda_{\mathcal{A}}}
\newcommand{\Lamt}{\Lambda_t}
\newcommand{\MA}{M_{\mathcal{A}}}

\newcommand{\MP}{M_{\mathcal{P}}}
\newcommand{\ML}{M_{\Lambda}}

\newcommand{\chiP}{\chi_{\mathcal{P}}}

\def\>{\rangle}
\def\<{\langle}

\begin{document}

\title{Modeling quantum noise for efficient testing of fault-tolerant circuits}

\author{Easwar Magesan}
\affiliation{Institute for Quantum Computing, University of Waterloo, Waterloo, ON N2L 3G1, Canada}
\affiliation{Department of Applied Mathematics, University of Waterloo, Waterloo, ON N2L 3G1, Canada}	
\author{Daniel Puzzuoli}
\affiliation{Institute for Quantum Computing, University of Waterloo, Waterloo, ON N2L 3G1, Canada}
\affiliation{Department of Applied Mathematics, University of Waterloo, Waterloo, ON N2L 3G1, Canada}	
\author{Christopher E. Granade}
\affiliation{Institute for Quantum Computing, University of Waterloo, Waterloo, ON N2L 3G1, Canada}
\affiliation{Department of Physics and Astronomy, University of Waterloo, Waterloo, ON N2L 3G1, Canada}	
\author{David G. Cory}
\affiliation{Institute for Quantum Computing, University of Waterloo, Waterloo, ON N2L 3G1, Canada}
\affiliation{Department of Chemistry, University of Waterloo, Waterloo, ON N2L 3G1, Canada}	
\affiliation{Perimeter Institute for Theoretical Physics, Waterloo, ON N2L 2Y5, Canada}

\begin{abstract}
Understanding fault-tolerant properties of quantum circuits is important for the design of large-scale quantum information processors. In particular, simulating properties of encoded circuits is a crucial tool for investigating the relationships between the noise model, encoding scheme, and threshold value. For general circuits and noise models, these simulations quickly become intractable in the size of the encoded circuit. We introduce methods for approximating a noise process by one which allows for efficient Monte Carlo simulation of properties of encoded circuits. The approximations are as close to the original process as possible without overestimating their ability to preserve quantum information, a key property for obtaining more honest estimates of threshold values. We numerically illustrate the method with various physically relevant noise models.
\end{abstract}

\maketitle


Representing and transforming information using the principles of quantum mechanics implies that a quantum information processor can solve certain problems exponentially faster than any currently known classical algorithm~\cite{Sho94,Fey82,Gro96}. Unfortunately, physically realizing such a processor is a difficult task as quantum systems are extremely sensitive to environmental noise effects. Hence, one will likely have to take advantage of quantum error-correction techniques~\cite{Sho95,CS,Ste96,KLL} and fault-tolerant encodings of quantum information~\cite{AB-O,KLZ,Pre97} to perform accurate large-scale quantum computation (QC). 

In order to verify fault-tolerant quantum computation is possible for a given noise model, one must prove that a ``threshold theorem" is satisfied. The main idea of such a theorem is that if the error-rate on the physical operations is below a threshold value $r_{th}$, and one can find an encoding that propagates errors in a controlled manner, then the error in the computation can be made arbitrarily small by concatenating the encoding. The 0-level corresponds to the physical unencoded operations, the 1-level corresponds to the first level of encoding and so on.

The ability to simulate the behaviour of encoded circuits by Monte Carlo methods can provide valuable information such as the existence and numerical estimation of $r_{th}$, as well as how errors propagate through encoded operations~\cite{Zal97,Ste03,AGP06,MM08,CDT09,Whi09}. In general, these simulations are inefficient on a classical computer for even moderately large systems, and so one typically makes assumptions about either the types of encoded gates included in the circuit or the noise model, or both. In particular, one often restricts attention to encoded stabilizer (Clifford) circuits~\cite{Got97}, which is not a significant limitation due to the ``magic state" model of QC~\cite{BK} where Clifford gates, ancilla magic states, and computational basis measurements suffice for universal QC. In certain cases, such as the seven-qubit Steane code~\cite{Ste96}, the encoded Clifford gates are comprised solely of 0-level Clifford gates and so encoded stabilizer circuits consist only of 0-level Clifford gates. By the Gottesman-Knill theorem~\cite{Got97} when these circuits are augmented with computational input states and measurements they are efficiently simulatable on a classical computer if noise is not taken into account. Unfortunately when noise is taken into consideration, stabilizer circuit simulations are generally no longer efficient. Hence, one often makes assumptions about the noise model at each faulty location, for instance, that the noise is described by a Pauli channel. More generally, when the noise at each fault location is modelled by a mixed-Clifford channel, classical Monte Carlo simulation is still possible by the Gottesman-Knill theorem. Pauli channels have various useful properties which include a simple geometric interpretation, closure under composition, and diagonal $\chi$ (process) matrices~\cite{CN} when expressed in the Pauli basis. As well, Pauli channels represent a wide class of physically realistic noise models such as dephasing and depolarizing processes. 

In reality, the noise at each location of the circuit will neither be a Pauli nor mixed-Clifford channel and so an important question for efficient simulation is how one can approximate the true noise at each location by one of these channels. One method is to diagonalize the noise in the Pauli basis by removing the off-diagonal elements of the $\chi$-matrix~\cite{CN}. This is an attractive method since in theory it can be performed experimentally via a procedure called Pauli twirling~\cite{SMKE}. Unfortunately, in practice, the twirling elements will not be implemented perfectly and exact diagonalization can not be achieved. Moreover, as we show later, twirling allows for the possibility that many states are better preserved under the twirled channel than the true channel, a scenario that is not ideal when attempting to find values of $r_{th}$.

The goal of this paper is to provide a method to approximate a noise process $\Lambda$ by a channel $\LamA$ such that:

\medskip

1. $\LamA$ is as ``close" (faithful) to $\Lambda$ as possible,

\medskip

2. $\LamA$ provides an ``honest" description of the reliability of $\Lambda$ for preserving quantum information,

\medskip

3. $\LamA$ allows for efficient simulations of properties of quantum circuits.

\medskip

Point 3, coupled with the Gottesman-Knill theorem, motivates analyzing the specific cases of $\LamA$ being either a Pauli or mixed-Clifford channel, which will be the focus of this paper. Ideally, these $\LamA$ will provide more realistic estimates of the threshold parameter. We emphasize however that the theory developed below holds for completely general approximations $\LamA$.

To quantify ``close" and ``honest" we will require rigorous methods for comparing quantum states and channels. Two standard methods for such comparisons are derived from the $1$-norm on linear operators, $\| \: \|_1$, and the diamond norm on linear superoperators, $\| \: \|_{\diamond}$~\cite{Kit97}. These comparative measures are ideal because they have a clear operational interpretation in terms of maximal distinguishability of quantum states (operations) via POVM measurements. 

Our goal can now be phrased precisely in the following manner:

\medskip

Suppose $\Lambda$ is a quantum channel whose complete description is given. We want to find the Pauli, or more generally mixed-Clifford, channel $\LamA$ that is the solution to the following constrained optimization problem:

\medskip

Minimize: $\|\LamA - \Lambda\|_{\diamond}$

\smallskip

Subject to: for every quantum state $\rho$, 
\begin{equation}
\|(\LamA - \mathcal{I} )(\rho)\|_{1} \geq \|(\Lambda - \mathcal{I} )(\rho)\|_{1}.\label{eq:Condition 1a}
\end{equation}

\noindent When $\LamA$ is a Pauli channel we denote it by $\LamP$ and when it is a mixed-Clifford channel it will be denoted by $\LamC$. For any quantum channel $\Lambda$ we call $\|\left(\Lambda - \mathcal{I}\right)(\rho)\|_1$ the ``input-output distinguishability" of $\Lambda$ with respect to $\rho$.

We first restrict attention to single-qubit, unital quantum channels and discuss generalizations to multi-qubit systems and non-unital channels later. Thus for now the Hilbert space of the quantum system is given by $\Hilb \equiv \mathbbm{C}^2$ and $\Lambda$ is assumed to map the maximally mixed state $\frac{\Id}{d}$ to itself. The single-qubit case is particularly relevant when the noise affecting the circuit is highly local. The validity of this assumption depends on many parameters such as the geometric layout of the circuit, the ability to address specific qubits involved in a particular operation~\cite{GCMJ}, and the form of the encoding. For instance, since generating sets of the unitary and Clifford group contain only one and two-qubit gates, it may be that the true noise is described by a highly local model when the encoded operations are performed transversally as in the Steane code. 


Our first task is to obtain a general state-independent form for Eq.~(\ref{eq:Condition 1a}). Afterwards we will discuss restrictions to our cases of interest. To begin, we make use of the Bloch sphere representation~\cite{Blo46} of single qubit states. For each state $\rho$ we can associate $\rho \rightarrow \vec{r}$ where the 3-vector $\vec{r}$ lies in the unit sphere of $\mathbbm{R}^3$. This vector is called the Bloch vector (representation) of $\rho$. For single qubit states, every point in the unit sphere is associated to a unique quantum state and the boundary (shell) of the unit sphere corresponds exactly to the set of pure states. 

Quantum channels also take a simple form in the Bloch sphere representation~\cite{BW,FA,RSW}. Any unital quantum channel $\Lambda$ can be uniquely represented by a real matrix $\ML$ such that $\vec{r} \rightarrow \ML \vec{r}$. This representation preserves many intuitive features of quantum operations, for instance a unitary operation $\mathcal{U}$ is represented by an orthogonal (rotation) matrix $M_{\mathcal{U}}$ and Pauli channels are represented by diagonal matrices.

We will exploit an extremely useful correspondence relating the $1$-norm distance between quantum states $\rho_1$, $\rho_2$ and the standard $2$-norm (Euclidean) distance between their Bloch sphere representations $\vec{r}_1$, $\vec{r}_2$, 
\begin{equation}
\|\rho_1-\rho_2\|_1 = \|\vec{r}_1 - \vec{r}_2\|_2. \label{eq:disteq}
\end{equation}
\noindent Eq.~(\ref{eq:disteq}) allows us to obtain a state-independent version of Eq.~(\ref{eq:Condition 1a}). Indeed, Eq~(\ref{eq:disteq}) implies Eq.~(\ref{eq:Condition 1a}) is equivalent to
\begin{equation}
\|\vec{r} - \MA \vec{r}\|_{2} \geq \|\vec{r} - \ML \vec{r} \|_{2} \label{eq:Condition 1aP}
\end{equation}
\noindent holding for all unit vectors $\vec{r}$. Using the theory of quadratic forms, Eq.~(\ref{eq:Condition 1aP}) is equivalent to $A\geq B$ where
\begin{eqnarray}
A &:=& \left(\Id-\MA\right)^T\left(\Id-\MA\right), \\
B &:=& \left(\Id-\ML\right)^T\left(\Id-\ML\right)
\end{eqnarray}
\noindent and ``$T$" denotes the transpose operation. In the case of a Pauli approximation $\LamP$ (which has a diagonal Bloch matrix representation), $A = \left(\Id-\MP\right)^2$. Since a description of $\Lambda$ is assumed to be given, $B$ can be computed in a straightforward manner~\cite{NC}. Thus finding values for the elements of $\MA$ which give $A-B \geq 0$ will ensure Eq.~(\ref{eq:Condition 1a}) is satisfied. Minimizing $\|\LamA - \Lambda\|_{\diamond}$ over these possible values gives the solution to our problem. 

We first look at Pauli channel approximations and then, using the intuition gained from these examples, discuss mixed Clifford channel approximations. Before analyzing the results in detail, let us set some notation. For a single qubit, the set $\{\sigma_0, \sigma_1, \sigma_2, \sigma_3\} = \{\Id,X,Y,Z\}$ is the usual orthogonal Pauli basis for the set of $2 \times 2$ complex matrices and is denoted by $\Paulone$. The extension to the multi-qubit Pauli basis $\Pauln$ is obtained by taking tensor products of elements of $\Paulone$. $\Pauln$ consists of traceless (except for $\Id$), unitary and Hermitian matrices and is a group when one includes phases. The Clifford group $\Clif$ is defined to be the normalizer of $\Pauln$ and can be generated by the Hadamard ($H$), phase ($S$) and CNOT gates applied on pairs of qubits. Note that $\Pauln$ is trivially contained in $\Clif$. A single-qubit mixed-Clifford channel $\LamC$ has the form $\LamC(\rho)=\sum_i p_iC_i\rho C_i^{\dagger}$ where the $p_i$ form a probability distribution, the $\{C_i\}$ form a subset of $\Clifone$, and $\rho$ is an arbitrary mixed state input to the channel. In the specific case of a Pauli channel $\LamP$ we have $\LamP(\rho) = \sum_{i=0}^{3} p_i \sigma_i\rho\sigma_i$.

 
\emph{Numerical Results - }\:  We perform the approximation scheme on three types of unital channels, each of which represents physically relevant noise: 
\begin{eqnarray}
\Lambda^{(1)}(\rho)&=&(1-p)\rho+p(\vec{n}_p\cdot\vec{\sigma})\rho(\vec{n}_p\cdot\vec{\sigma}),\\
\Lambda^{(2)}(\rho)&=&(1-3p)\rho+p\sum_{i=1}^3\sigma_i\rho \sigma_i,\\
\Lambda^{(3,k)}(\rho)&=&\exp\left(-i\frac{\theta}{2}\vec{n}_k\cdot\vec{\sigma}\right)\rho \exp\left(i\frac{\theta}{2}\vec{n}_k\cdot\vec{\sigma}\right).
\end{eqnarray}
The parameter values we choose are $p=0.01$, $\vec{n}_p = (\sin(\pi/8), 0, \cos(\pi/8))$, $\theta=0.02$, and $\vec{n}_k = (\sin(\theta_k), 0, \cos(\theta_k))$ where $\theta_k=\frac{k\pi}{8}$ for $k\in\{0,1,2,3,4\}$. $\Lambda^{(1)}(\rho)$ represents dephasing noise about a non-Pauli axis, $\Lambda^{(2)}(\rho)$ represents a depolarizing channel and the $\Lambda^{(3,k)}(\rho)$ represent rotations about axes in the $x-z$ plane starting from the $z$ axis and ending at the $x$-axis. The depolarizing channel was included in the analysis to verify that when $\Lambda$ is itself a Pauli channel, the scheme returns $\LamP = \Lambda$. The Pauli approximations $\LamP$ for each case are given in Table \ref{table:numericalresults1} via their $\chi$-matrix $\chiP$. The diamond norm distance for each case is also given and was calculated using the semidefinite program of Ref.~\cite{Wat09}.

\begin{table}[h!]
	\caption{\label{table:numericalresults1}Pauli channel approximations $\LamP^{(i)}$ and diamond norm distance between $\LamP^{(i)}$ and $\Lambda^{(i)}$.}
	\begin{tabular}{ccccccc}
		\toprule
		Channel		& \multicolumn{4}{c}{Approximation $\LamP^{(i)}$} 					              &$\|\LamP^{(i)}-\Lambda^{(i)}\|_{\diamond}$                   \\
					& $\left[\chiP\right]_{0,0}$	& $\left[\chiP\right]_{1,1}$	& $\left[\chiP\right]_{2,2}$	& $\left[\chiP\right]_{3,3}$	        &                                                                                                                                                           \\
					\hline
		$\Lambda^{(1)}$	&	0.9860	&	0.0020	&	0.0040	&	0.0080	&	0.0152                                                                                                                     \\
		$\Lambda^{(2)}$	&	0.9700	&	0.0100	&	0.0100	&	0.0100	&	0                                                                                                               \\
		$\Lambda^{(3,0)}$	&	0.9900	&	0	&	0	&	0.0100	&	0.0281                                                                                                                 \\
		$\Lambda^{(3,1)}$	&	0.9860	&	0.0022	&	0.0040	&	0.0078	&	0.0359                                                                                                                \\
		$\Lambda^{(3,2)}$	&	0.9850	&	0.0050	&	0.0050	&	0.0050	&	0.0381                                                                                                                  \\
		$\Lambda^{(3,3)}$	&	0.9860	&	0.0078	&	0.0040	&	0.0022	&	0.0359                                                                                                                  \\
		$\Lambda^{(3,4)}$	&	0.9900	&	0.0100	&	0	&	0	&	0.0281                                                                                                                 \\
		\hline

	\end{tabular}
	
\end{table}
We give $\chi_{\mathcal{P}}$ rather than $\MP$ because each can be obtained from the other in a straightforward manner~\cite{NC}, the $\chi$-matrix is a standard tool in process tomography~\cite{CN}, and any entry of the $\chi$-matrix can be directly estimated via experiments~\cite{ML06,KLRB,ESMR,MGE,MGR,BPP,SKMR}. In particular, for any quantum channel $\Lambda$, $\chi_{0,0}$ is directly related to the average fidelity of $\Lambda$, $\overline{\mathcal{F}_{\Lambda,\mathcal{I}}} = \int \tr\left(\Lambda(\kett\braa)\kett\braa\right) d\psi$, which is a standard experimental figure of merit for how close an intended unitary operation is to the implemented operation. The $\chi_{0,0}$ elements of $\Lambda^{(1)}$, $\Lambda^{(2)}$ and the $\Lambda^{(3,j)}$ are given in Table~\ref{table:numericalresults0} below.
\begin{table}[h!]
	\caption{\label{table:numericalresults0}$\chi_{0,0}$ elements of $\Lambda^{(1)}$, $\Lambda^{(2)}$, $\Lambda^{(3,j)}$ for $j=0,1,2$.}
	\begin{tabular}{cccc}
		\toprule
		Channel:		& $\Lambda^{(1)}$ & $\Lambda^{(2)}$ & $\Lambda^{(3,j)}$				                            \\
		$\chi_{0,0}$:	&	0.9900	     &	0.9700                       &	0.9999                                                                                    \\
		\hline

	\end{tabular}
	\end{table}
As expected from the constraint of Eq.~(\ref{eq:Condition 1a}), $[\chiP]_{0,0}^{(i)} \leq \chi_{0,0}^{(i)}$ in all cases. As well, since $\Lambda^{(2)}$ is depolarizing, $\LamP^{(2)}=\Lambda^{(2)}$ which verifies our consistency check. For the unitary rotations $\Lambda^{(3,j)}$, the decrease in average fidelity is relatively large and is a maximum at $k=2$. When $k=0$, $\LamP^{(3,0)}$ is dephasing about the $z$-axis, and as the rotation axis angle approaches $\frac{\pi}{4}$, the approximation converges to a depolarizing channel. By symmetry, as the rotation axis angle goes to $\frac{\pi}{2}$, $\LamP^{(3,4)}$ becomes dephasing.

It is straightforward to show there is a dephasing channel which exactly reproduces the input-output distinguishability of a unitary rotation about any axis. Indeed, for every rotation angle $\theta$ and state $\rho$ whose Bloch vector $\vec{r}$ is at an angle $\alpha$ relative to $z$, 
\begin{equation}
\|(\Lambda^{(3,0)} - \mathcal{I} )(\rho)\|_{1} = 2\left|\sin\left(\frac{\theta}{2}\right)\right||\sin(\alpha)| \|\vec{r}\|_2.
\end{equation}
As well, for a dephasing channel $\LamD^{(3,0)}$ about $z$ given by $\LamD^{(3,0)}(\rho)=(1-p)\rho+p\sigma_z\rho\sigma_z$, the input output distinguishability has a very similar form: 
\begin{equation}
\|(\LamD^{(3,0)} - \mathcal{I} )(\rho)\|_{1} = 2p|\sin(\alpha)|\|\vec{r}\|_2.
\end{equation}
Hence, setting $p=|\sin(\theta/2)|$ implies one can \emph{exactly} match the input-output distinguishability of $\Lambda^{(3,0)}$ by a dephasing channel. Interestingly, this dephasing channel is the optimal channel found by our algorithm. More generally, a rotation about any axis $\hat{n}$ and a dephasing channel about $\hat{n}$ have the same input-output distinguishability if $p=|\sin(\theta/2)|$ and, if the rotation is about a Pauli axis, this dephasing channel is a Pauli channel.

\begin{figure}[h!]
\centering
\includegraphics[width=85mm,height=115mm]{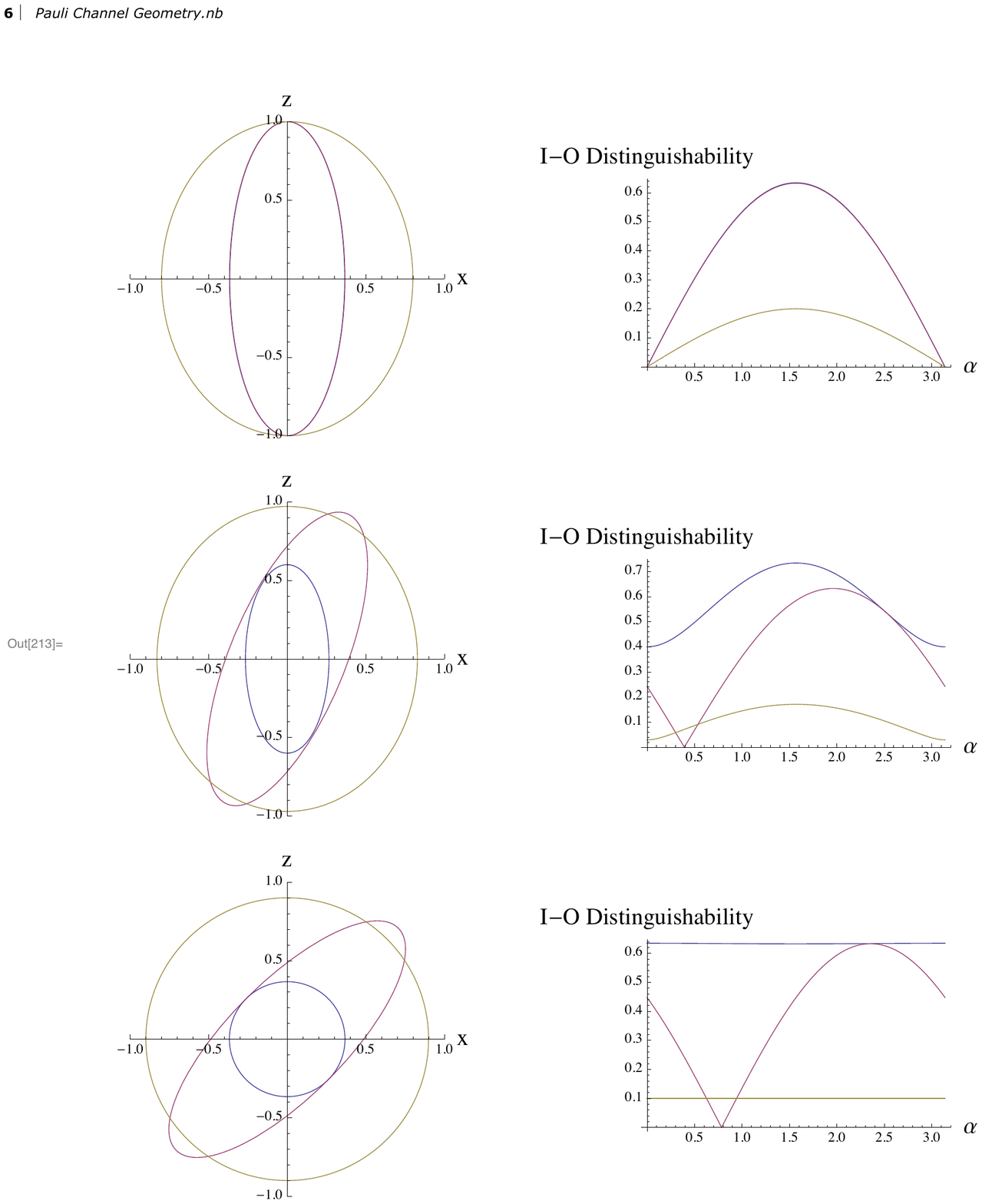}
\caption{\label{fig:1} $\LamP^{(3,j)}$ (blue), $\LamD^{(3,j)}$ (red), $\Lamt^{(3,j)}$ (gold) for $j=0,1,2$. (Left Column): Action on the x-z plane of the Bloch sphere. (Right Column): Input-Output distinguishability for states $\rho$ at angle $\alpha$ relative to $z$-axis.}
\end{figure}

For each $j$, we can use the Bloch sphere to visualize the difference between $\LamP^{(3,j)}$ and the dephasing channel described above which exactly matches the input-output distinguishability for each $\rho$. We denote these dephasing channels by $\LamD^{(3,j)}$. Fig.~\ref{fig:1} contains plots of the deformation of the Bloch sphere in the x-z plane by $\LamP^{(3,j)}$ (blue) and $\LamD^{(3,j)}$ (red) for $j=0,1,2$. To make the visualization more apparent we rotate by a larger angle, $\theta=2\sin^{-1}(\sqrt{0.1})$.

Clearly $\LamP^{(3,j)}$ diverges from $\LamD^{(3,j)}$ as $\theta_j$ goes to $\frac{\pi}{4}$. Hence, the best Pauli approximation becomes significantly worse than the best dephasing approximation. This indicates it can be useful to search over more general classes of channels than Pauli channels. For instance if we augment $\Paulone$ with the Hadamard gate $H$ then at $j=2$ our approximation $\LamA$ is a mixed-Clifford channel and is exactly $\LamD^{(3,2)}$. This is because $\LamD^{(3,2)}$ has the form
\begin{equation}
\LamD^{(3,2)}(\rho)=(1-p)\rho + pH\rho H
\end{equation}
where $p=|\sin(\theta /2)| = \sqrt{0.1}$. Thus knowing the form of the noise $\Lambda$ can provide intuition as to what gates can be augmented to $\Paulone$ to obtain a better approximation.

Another example helps to illustrate this point. Let us look again at approximating $\Lam^{(3,0)}$, however this time we augment the Pauli group with the Clifford operator $Z_{\pi /2}=\exp\left(-i\frac{\pi}{4}\sigma_z\right)$. Denoting the approximation by $\Lambda_{Z_{\pi /2}}$ we obtain the results in Table~\ref{table:Z90} below (the results for using only Pauli operators are in Table~\ref{table:numericalresults1}).
\begin{table}[h!]
	\caption{\label{table:Z90}  Parameters for Approximation $\Lambda_{Z_{\pi /2}}$.}
	\begin{tabular}{ccccccc}
		\toprule
		Channel	\: \:	& \multicolumn{4}{c}{Approximation $\Lambda_{Z_{\pi /2}}$} 				\: \:	              &$\|\Lambda_{Z_{\pi /2}}-\Lambda^{(3,0)}\|_{\diamond}$                   \\
					& $\chi_{00}$	& $\chi_{11}$	& $\chi_{22}$	& $\chi_{33}$	        &                                                                                                                                                           \\
					\hline
		$\Lambda^{(3,0)}$	&	0.9929	&	0	&	0	&	0.0071	&	0.0151                                                                                                                 \\
		\hline

	\end{tabular}
	
\end{table}
The average fidelity increased and $\|\Lambda_{Z_{\pi /2}}-\Lambda^{(3,0)}\|_{\diamond}$ is significantly smaller than $\|\LamP^{(3,0)}-\Lambda^{(3,0)}\|_{\diamond}$. Both of these results are expected since $Z_{\frac{\pi}{2}}$ provides better information for the approximation than $Z_{\pi}$.


Let us now compare some of the Pauli approximations of Table~\ref{table:numericalresults1} with those that would be obtained from twirling $\Lambda$ over the Pauli group. For an arbitrary channel $\Lambda$, we denote the Pauli twirl of $\Lambda$ by $\Lambda_t$. We analyze how $\Lambda_t^{(i)}$ and $\LamP^{(i)}$ compare in terms of the two conditions we require for an optimal Pauli approximation: first, how $\|\LamP^{(i)}-\mathcal{I}\|_{\diamond}$ compares to $\|\Lambda_t^{(i)}-\mathcal{I}\|_{\diamond}$ for $i = 1,(3,0),(3,1),(3,2)$ (see Table~\ref{table:trunc vs approx2}), and second, how $\|\left(\LamP^{(3,j)}-\mathcal{I}\right)(\rho)\|_1$ and $\|\left(\Lambda_t^{(3,j)}-\mathcal{I}\right)(\rho)\|_1$ compare for $j=0,1,2$ and states $\rho$ at angle $\alpha$ relative to the $z$-axis (see right hand side of Fig.~\ref{fig:1}).

We see that for all channels examined $\|\Lambda^{(i)}-\Lambda_t^{(i)}\|_{\diamond}\leq\|\Lambda^{(i)}-\LamP^{(i)}\|_{\diamond}$ and so the Pauli twirl is a better model of the original channel. However from the right-hand column of Fig.~\ref{fig:1} we see that, except for a small set of states, $\|(\LamP^{(i)}-\mathcal{I})(\rho)\|_1\geq\|(\Lambda^{(i)}-\mathcal{I})(\rho)\|_1\geq\|(\Lambda_t^{(i)}-\mathcal{I})(\rho)\|_1$ which demonstrates that the truncation underestimates the error and our channel over-estimates as designed. So, for the price of a slightly larger diamond norm distance to the true error channel $\Lambda$, we have ensured that we do not underestimate the error. This can be crucial when considering estimates of the threshold.
\begin{table}[tp]
	\caption{\label{table:trunc vs approx2}Comparisons of $\|\Lambda^{(i)} - \Lamt^{(i)}\|_{\diamond}$ vs $\|\Lambda^{(i)} - \LamP^{(i)}\|_{\diamond}$.}
	\begin{tabular}{cccc}
		\toprule
		Channel		\:&	$\|\Lambda^{(i)}-\Lambda_t^{(i)}\|_{\diamond}$	  \: &	$\|\Lambda^{(i)}-\LamP^{(i)}\|_{\diamond}$	\\
					\hline
		$\Lambda^{(1)}$	&	0.0071	&	0.0152	\\
		$\Lambda^{(3,0)}$	&	0.0020	&	0.0281	\\
		$\Lambda^{(3,1)}$	&	0.0020	&	0.0359	\\
		$\Lambda^{(3,2)}$	&	0.0020	&	0.0381	\\
		\hline

	\end{tabular}
	
\end{table}


We now discuss generalizations of these results to both the non-unital and multi-qubit cases. The Bloch sphere representation of a non-unital single-qubit channel $\Lambda$ is completely specified by a matrix $\ML$ and a vector $\vec{t}$ which represents the non-unitality of the map,
\begin{equation}
\vec{r} \mapsto \ML\vec{r} + \vec{t}.
\end{equation}
One can show using a similar argument with quadratic forms that if $A\geq B$ where
\begin{eqnarray}
A &:=& \left(\Id-\MA\right)^T\left(\Id-\MA\right), \\
B &:=& \left(\Id-\ML\right)^T\left(\Id-\ML\right) + \left(\|\vec{t}\|_2^2 + 2\|\vec{v}\|_2\right)\Id
\end{eqnarray}
then for every $\vec{r}$,
\begin{equation}
\|\vec{r} - \MA \vec{r}\|_{2} \geq \|\vec{r} - \left(\ML \vec{r} + \vec{t}\right) \|_{2}. \label{eq:Condition 1anu}
\end{equation}
Hence for every state $\rho$, Eq.~(\ref{eq:Condition 1a}) is satisfied where the vector $\vec{v}$ above is given by $\vec{v}=\left(\Id - \ML\right)^T\vec{t}$.


For the multi-qubit case, it is not true in general that for states $\rho_1$ and $\rho_2$, $\|\rho_1-\rho_2\|_1 = \|\vec{r_1}-\vec{r_2}\|_2$, however it is still likely the case that if $A \geq B$ then Eq.~(\ref{eq:Condition 1a}) is satisfied. The multi-qubit case is of significance when considering correlated noise models in encoded circuits. For instance, while many treatments of noise models in fault-tolerant circuits assume local, stochastic noise models, it is entirely possible that errors at certain locations can imply errors occur at other specific locations. It can also be the case that two locations always feel the same environmental influence and thus will undergo collective noise. Here we numerically analyze the collective unitary noise model $\Lambda^{(2q)}$ with Kraus operator $\exp\left(-i0.01 \sigma_X\otimes \sigma_X\right)$ (a two-qubit rotation about the x-axis by $\theta = 0.02$). The $\chi$-matrix of this channel is $16\times 16$ but, given that it only contains $\Id\otimes \Id$ and $\sigma_x \otimes \sigma_x$ terms, it is sparse and can be represented by a $2\times 2$ matrix. 
\begin{eqnarray}
	\chi^{(2q)}&=&
	\left[ \begin{array}{cc}
	0.9999	&	0.0100i\\
	-0.0100i 	&	0.0001
	\end{array} \right].
\end{eqnarray}
The Pauli channel approximation for this channel, $\LamP^{(2q)}$, is displayed in Table \ref{table:2qubit}. As expected, $\LamP^{(3,4)}$ is symmetric across both qubits and interestingly gives the exact same results as for the single-qubit rotation about $\sigma_X$, $\LamP^{(3,4)}$, given in Table~\ref{table:numericalresults1}.
\begin{table}[tp]
	\caption{\label{table:2qubit} Pauli channel approximation details for $\Lambda^{(2q)}$.}
	\begin{tabular}{cccc}
		\toprule
		Channel	\:\:	& \multicolumn{2}{c}{Approximation $\LamP^{(2q)}$} 	\: \:				              &$\|\LamP^{(2q)}-\Lambda^{(2q)}\|_{\diamond}$                     \\
					& $\chi_{00}$	& $\chi_{xx}$                                                                                                                                                          \\
					\hline
		$\Lambda^{(2q)}$	&	0.9900	&	0.0100	  & 0.0281                                                                                                                  \\
	\hline
		
	\end{tabular}
	
\end{table}

To conclude, we have provided a method for approximating quantum operations such that the approximations are as close to the true operation as possible without overestimating its ability to preserve quantum information. We have explicitly analyzed single-qubit Pauli approximations, discussed how one can extend the analysis to mixed-Clifford approximations, and have shown that alternative methods such as twirling the original channel leads to highly dishonest approximations. We have also presented a rigorous extension to the non-unital case and have provided numerical evidence that our method likely holds in the multi-qubit case as well. These results are essential for simulating quantum circuits as they allow for both more honest estimates of threshold values as well as scalable simulation of circuit properties such as error propagation.

We acknowledge helpful discussions with Joseph Emerson, Daniel Gottesman, Ian Hincks, and Marcus Silva. The authors acknowledge support from NSERC, CIFAR, CERC, and the Ontario government. This work was partially 
supported by the Intelligence Advanced Research Projects Activity (IARPA) via Department of Interior National Business Center Contract number DllPC20l66. The U.S. Government is authorized to reproduce and distribute reprints for Governmental purposes notwithstanding any copyright annotation thereon. Disclaimer: The views and conclusions contained herein are those of the authors and should not be interpreted as necessarily representing the official policies or endorsements, either expressed or implied, of IARPA, DoI/NBC or the U.S. Government.

%

\end{document}